\documentclass{cimento}
\usepackage{graphicx}
\usepackage{cite}
\usepackage{amsmath}
\usepackage{amssymb}
\usepackage{bm}
\usepackage[T1]{fontenc}
\usepackage{mathrsfs}
\usepackage{dcolumn}

\def\ve#1{{\bm{#1}}}
\def\nuc#1#2#3{{}^{#2}_{#3}\mathrm{#1}}
\def\urm#1{\scriptstyle{\text{\textrm{\textmd{\textup{#1}}}}}}

\def\ca#1{{\mathcal{#1}}}

\let\temp\epsilon
\let\epsilon\varepsilon
\let\varepsilon\temp
\let\temp\relax
\let\temp\phi
\let\phi\varphi
\let\varphi\temp
\let\temp\relax
\DeclareMathOperator{\laplace}{\Delta}

\title{Possible inconsistency between phenomenological and theoretical determinations of charge symmetry breaking in nuclear energy density functionals}
\shorttitle{Possible inconsistency of charge symmetry breaking in nuclear EDF}
\author{Tomoya~Naito\from{ins:ithems}\from{ins:ut},
  Gianluca~Col\`{o}\from{ins:um}\from{ins:infn},
  Tetsuo~Hatsuda\from{ins:ithems},
  Haozhao~Liang\from{ins:ut}\from{ins:ithems},
  Xavier~Roca-Maza\from{ins:um}\from{ins:infn},
  \atque
  Hiroyuki~Sagawa\from{ins:aizu}\from{ins:rnc}}
\shortauthor{T.~Naito, G.~Col\`{o}, T.~Hatsuda, H.~Liang, X.~Roca-Maza, \atque H.~Sagawa}
\instlist{
  \inst{ins:ithems}
  RIKEN Interdisciplinary Theoretical and Mathematical Sciences Program, Wako, Japan
  \inst{ins:ut}
  Department of Physics, The University of Tokyo,
  Tokyo, Japan
  \inst{ins:um}
  Dipartimento di Fisica, Universit\`{a} degli Studi di Milano,
  Milano, Italy
  \inst{ins:infn}
  INFN, Sezione di Milano,
  Milano, Italy
  \inst{ins:aizu}
  Center for Mathematics and Physics, University of Aizu,
  Aizu-Wakamatsu, Japan
  \inst{ins:rnc}
  RIKEN Nishina Center, Wako, Japan}
\begin{document}
\maketitle
\begin{abstract}
  We summarize the recent progress on the determination of the charge symmetry breaking term of nuclear energy density functionals.
  We point out that the strength of the term determined theoretically is remarkably smaller than that determined phenomenologically,
  which is still an open question.
\end{abstract}
%
\section{Introduction}
\par
The idea of the isospin $ T $ was introduced in the 1930s to distinguish protons and neutrons as two different states of nucleons ~\cite{
  Heisenberg1932Z.Phys.77_1,
  Wigner1937Phys.Rev.51_106}.
The isospin symmetry of the nuclear interaction was also proposed in the 1930s~\cite{
  Cassen1936Phys.Rev.50_846,
  Kemmer1938Math.Proc.Camb.Philos.Soc.34_354},
i.e., the nuclear interaction does not depend on the third component of the isospin $ T_z $.
Accordingly, the proton-proton, the neutron-neutron, and the $ T = 1 $ channel of the proton-neutron channels are identical to each other.
It is known that this symmetry is, however, just an approximate one and slightly broken.
This is evidenced by the difference in the scattering length.
The main isospin symmetry breaking (ISB) terms of the nuclear interaction can be classified into two:
charge independence breaking (CIB)
and 
charge symmetry breaking (CSB).
They are also, respectively, referred to as class II and III nuclear interactions~\cite{
  Henley1979MesonsinNucleiVolumeI_405}.
The former corresponds to
the difference between like-particle interaction and different-particle interaction; 
the latter corresponds to
the difference between proton-proton interaction and neutron-neutron one.
\par
Isospin symmetry of atomic nuclei is also broken due to the Coulomb interaction.
Okamoto~\cite{
  Okamoto1964Phys.Lett.11_150}
and
Nolen and Schiffer~\cite{  
  Nolen1969Annu.Rev.Nucl.Sci.19_471}
pointed out that the Coulomb interaction is not enough to describe the mass difference of mirror nuclei,
a pair of two nuclei one of which is composed of $ Z $ protons and $ N $ neutrons
and the other of which is of $ N $ protons and $ Z $ neutrons.
Recently, thanks to the progress of nuclear experiments,
the shape and the spin-parity of mirror nuclei have been also discussed~\cite{
  Hoff2020Nature580_52,
  Wimmer2021Phys.Rev.Lett.126_072501}.
\par
Some bare nuclear interactions contain the ISB part~\cite{
  Wiringa1995Phys.Rev.C51_38,
  Machleidt2001Phys.Rev.C63_024001}.
Accordingly, effects of the ISB terms on nuclear properties have also been discussed in \textit{ab initio} calculations~\cite{
  Wiringa2013Phys.Rev.C88_044333,
  Novario2023Phys.Rev.Lett.130_032501}.
Nevertheless, only a few calculations of mean-field or density functional theory 
have considered the ISB terms of nuclear interaction~\cite{
  Roca-Maza2018Phys.Rev.Lett.120_202501,
  Baczyk2018Phys.Lett.B778_178,
  Baczyk2019J.Phys.G46_03LT01},
while such calculations are needed for a systematical study of effects of the ISB terms on nuclear properties.
In order to perform such a systematical study,
first, the effective interaction or energy density functional (EDF) for the ISB nuclear interaction should be determined.
In this proceeding, we summarize the recent progress on the determination of
an EDF of the CSB nuclear interaction,
since the CSB interaction dominates for most properties on the nuclear ground state~\cite{
  Naito2023Phys.Rev.C107_064302}.
%
%
%
\section{Skyrme-like isospin symmetry breaking interaction}
\par
To perform the Hartree-Fock calculation, one needs to introduce an effective nuclear interaction or an EDF.
We use a Skyrme Hartree-Fock calculation~\cite{
  Vautherin1972Phys.Rev.C5_626},
and accordingly, we also introduce the Skyrme-like zero-range CSB interaction as follows:
\begin{align}
  v_{\urm{Sky}}^{\urm{CSB}} \left( \ve{r} \right)
  & =
    \left\{
    s_0
    \left( 1 + y_0 P_{\sigma} \right)
    \delta \left( \ve{r} \right)
    \vphantom{
    +
    \frac{s_1}{2}
    \left( 1 + y_1 P_{\sigma} \right)
    \left[
    \ve{k}^{\dagger 2} \delta \left( \ve{r} \right)
    +
    \delta \left( \ve{r} \right) \ve{k}^2
    \right]
    +
    s_2
    \left( 1 + y_2 P_{\sigma} \right)
    \ve{k}^{\dagger}
    \cdot
    \delta \left( \ve{r} \right)
    \ve{k}}
    \right.
    \notag \\
  & \quad
    \left.
    +
    \frac{s_1}{2}
    \left( 1 + y_1 P_{\sigma} \right)
    \left[
    \ve{k}^{\dagger 2} \delta \left( \ve{r} \right)
    +
    \delta \left( \ve{r} \right) \ve{k}^2
    \right]
    +
    s_2
    \left( 1 + y_2 P_{\sigma} \right)
    \ve{k}^{\dagger}
    \cdot
    \delta \left( \ve{r} \right)
    \ve{k}
    \right\}
    \times
    \frac{\tau_{z1} + \tau_{z2}}{4},
    \label{eq:Skyrme_CSB} 
\end{align}
where
$ P_{\sigma} = \left( 1 + \ve{\sigma}_1 \cdot \ve{\sigma}_2 \right) / 2 $ is the spin-exchange operator,
$ \ve{k} $ is the relative momentum,
$ \tau_{zi} $ is the $ z $-projection of the isospin operator $ \ve{\tau} $ for the particle $ i $,
and
$ s_i $ and $ y_i $ ($ i = 0 $, $ 1 $, and $ 2 $) are parameters to be determined.
Accordingly, the energy density reads
\begin{align}
  \ca{E}_{\urm{CSB}} \left( \rho_p, \rho_n \right)
  & = 
    \frac{s_0}{8}
    \left( 1 - y_0 \right)
    \left( \rho_n^2 - \rho_p^2 \right)
    +
    \frac{1}{16}
    \left[
    s_1 \left( 1 - y_1 \right)
    +
    3 s_2 \left( 1 + y_2 \right)
    \right]
    \left( \rho_n t_n - \rho_p t_p \right)
    \notag \\
  & \quad
    -
    \frac{1}{64}
    \left[
    9 s_1 \left( 1 + y_1 \right)
    -
    s_2 \left( 1 - y_2 \right)
    \right]
    \left( \rho_n \laplace \rho_n - \rho_p \laplace \rho_p \right)
    \notag \\
  & \quad
    -
    \frac{1}{32}
    \left[
    s_1 \left( y_1 - 1 \right)
    +
    s_2 \left( y_2 + 1 \right) 
    \right]
    \left( \ve{J}_n^2 - \ve{J}_p^2 \right),
    \label{eq:CSB_tot} 
\end{align}
where $ \rho_{\tau} $, $ t_{\tau} $, and $ \ve{J}_{\tau} $ are the particle, kinetic, and the spin-orbit current densities for nucleon $ \tau $ ($ \tau = p $, $ n $), respectively.
Hereinafter, new parameters
\begin{equation}
  \tilde{s}_0
  \equiv
  s_0 \left( 1 - y_0 \right),
  \quad
  \tilde{s}_1
  \equiv
  s_1 \left( 1 - y_1 \right),
  \quad
  \tilde{s}_2
  \equiv
  s_2 \left( 1 + y_2 \right)
\end{equation}
are used for simplicity,
instead of the original $ s_j $ and $ y_j $ ($ j = 0 $, $ 1 $, and $ 2 $).
Note that
Eq.~\eqref{eq:CSB_tot} for the homogeneous systems can be written only by $ \tilde{s}_j $ 
\begin{equation}
  \label{eq:CSB_LDA}
  \ca{E}_{\urm{CSB}} \left( \rho_p, \rho_n \right)
  =
  \frac{\tilde{s}_0}{8}
  \left( \rho_n^2 - \rho_p^2 \right)
  +
  \frac{1}{16}
  \left(
    \tilde{s}_1 + 3 \tilde{s}_2 
  \right)
  \left( \rho_n t_n - \rho_p t_p \right)
\end{equation}
since $ \laplace \rho $ and $ \ve{J} $ terms vanish.
%
%
%
\section{Strength of Skyrme-like CSB interactions}
\par
To perform calculations, the parameters of Eq.~\eqref{eq:Skyrme_CSB} must be determined.
One way to determine these parameters is a fit to experimental data
as done in the isospin symmetric part of most EDFs,
hereinafter called ``phenomenological'' determination.
The other way to pin down the strengths is based on \textit{ab initio} calculations, if available, or even on quantum chromodynamics (QCD).
This section is devoted to summarizing the results of these two methods.
\subsection{Phenomenological determination}
\par
The phenomenological determination can be further classified into two:
in one, all the parameters, including the isospin symmetric part, are fitted to experimental data altogether,
while 
in the other, only the parameters of the ISB part are fitted independently on top of an effective interaction or EDF already existing.
Note that it is discussed in Ref.~\cite{
  Naito2022Phys.Rev.C105_L021304}
that these two methods may differ by $ 10 \, \% $ of the value of $ s_0 $.
\par
An EDF named ``SAMi-ISB''~\cite{
  Roca-Maza2018Phys.Rev.Lett.120_202501},
which includes only the leading-order ISB term ($ s_0 $ term),
is classified into the former type,
where $ y_0 = -1 $ is fixed to select the spin-singlet term.
The isobaric analog energy of $ \nuc{Pb}{208}{} $ and the energy difference of symmetric nuclear matter calculated with and without ISB terms,
as well as the criteria used in the original SAMi EDF,
are used to determine the parameters.
\par
Other EDFs, ``SLy4-ISB'', ``SkM*-ISB'', and ``$ \text{SV}_{\text{T}} $-ISB''~\cite{
  Baczyk2018Phys.Lett.B778_178},
are classified into the latter type.
In these EDFs, the parameters of the isospin symmetric terms remain those of the original form,
e.g., SLy4.
On top of an existing EDF, the parameter of the leading-order ISB term is fitted to experimental data
of mass differences of mirror nuclei,
which are sensitive to the Coulomb and CSB interactions~\cite{
  Baczyk2018Phys.Lett.B778_178,
  Naito2023Phys.Rev.C107_064302},
with $ y_0 = 0 $.
The ISB interaction with the next-leading-order ISB ($ s_0 $--$ s_2 $) terms on top of the $ \text{SV}_{\text{T}} $ interaction
with $ y_j = 0 $ ($ j = 0 $, $ 1 $, and $ 2 $)
are also constructed with the same methods~\cite{
  Baczyk2019J.Phys.G46_03LT01}.
\par
The other method proposed in Ref.~\cite{
  Sagawa2022Phys.Lett.B829_137072}
uses the isovector density
$ \rho_{\urm{IV}} \left( \ve{r} \right) = \rho_n \left( \ve{r} \right) - \rho_p \left( \ve{r} \right) $,
where $ \rho_n $ and $ \rho_p $ are, respectively, neutron and proton densities,
of $ \nuc{Ca}{40}{} $.
It was found in Ref.~\cite{
  Sagawa2022Phys.Lett.B829_137072}
that the peak height of $ \rho_{\urm{IV}} \left( r \right) $ is proportional to the CSB strength $ s_0 $
if only the $ s_0 $-term is considered.
On top of the isospin-symmetric part of the SAMi-ISB interaction, the CSB strength is re-determined.
\par
The strengths determined in these methods are listed in Table~\ref{tab:CSB}.
It can be found that the strength $ \tilde{s}_0 $ ranges from about $ -20 $ to $ -50 \, \mathrm{MeV} \, \mathrm{fm}^3 $ 
if one limits oneself to the leading-order interaction.
However, there are factor two differences between $ \tilde{s}_0 $ of the SAMi-ISB EDF and those of the others.
If the higher-order ($ \tilde{s}_1 $ and $ \tilde{s}_2 $) terms are included,
even the sign of $ \tilde{s}_0 $ is opposite from that determined without the $ \tilde{s}_1 $ or $ \tilde{s}_2 $ term.
This point will be discussed in detail later.

\begin{table}[tb]
  \centering
  \caption{Strengths of the various Skyrme-like CSB interactions.
    ``Pheno'' and ``Theor'', respectively, refer to results based on phenomenological fits and on theoretical evaluation.
    The values shown here except \textit{ab initio} determinations are taken from Refs.~\cite{
      Roca-Maza2018Phys.Rev.Lett.120_202501,
      Baczyk2018Phys.Lett.B778_178,
      Baczyk2019J.Phys.G46_03LT01,
      Sagawa2022Phys.Lett.B829_137072,
      Sagawa:2023itk}.}
  \label{tab:CSB}
  \begin{tabular}{lllll}
    \hline 
    Class & Method or Name & \multicolumn{1}{c}{$ \tilde{s}_0 $ ($ \mathrm{MeV} \, \mathrm{fm}^3 $)} & \multicolumn{1}{c}{$ \tilde{s}_1 $ ($ \mathrm{MeV} \, \mathrm{fm}^5 $)} & \multicolumn{1}{c}{$ \tilde{s}_2 $ ($ \mathrm{MeV} \, \mathrm{fm}^5 $)} \\
    \hline
    Pheno & SAMi-ISB                                          & $ -52.6 \pm  1.4 $ & \multicolumn{1}{c}{---} & \multicolumn{1}{c}{---} \\
    Pheno & SLy4-ISB (leading order)                          & $ -22.4 \pm  4.4 $ & \multicolumn{1}{c}{---} & \multicolumn{1}{c}{---} \\
    Pheno & SkM*-ISB (leading order)                          & $ -22.4 \pm  5.6 $ & \multicolumn{1}{c}{---} & \multicolumn{1}{c}{---} \\
    Pheno & $ \text{SV}_{\text{T}} $-ISB (leading order)      & $ -29.6 \pm  7.6 $ & \multicolumn{1}{c}{---} & \multicolumn{1}{c}{---} \\
    Pheno & $ \text{SV}_{\text{T}} $-ISB (next-leading order) & $ +44   \pm  8   $ & $ -56 \pm 16 $ & $ -31.2 \pm 3.2 $ \\
    Pheno & Estimation by isovector density                   & $ -17.6 \pm 32.0 $ &  \multicolumn{1}{c}{---} & \multicolumn{1}{c}{---} \\
    \hline
    Theor & $ \Delta E_{\urm{tot}} $ ($ \text{N}^2 \text{LO}_{\text{GO}} $ (394) \& CC) & $ -4.2 \pm 6.5 $ &  \multicolumn{1}{c}{---} & \multicolumn{1}{c}{---} \\
    Theor & $ \Delta E_{\urm{tot}} $ ($ \text{N}^2 \text{LO}_{\text{GO}} $ (450) \& CC) & $ -5.1 \pm 28.5 $ &  \multicolumn{1}{c}{---} & \multicolumn{1}{c}{---} \\
    Theor & $ \Delta E_{\urm{tot}} $ (AV18-UX \& GFMC)                                  & $ -6.413 \pm 0.173 $ &  \multicolumn{1}{c}{---} & \multicolumn{1}{c}{---} \\
    Theor & QCD sum rule (Case I)                                                       & $ -15.5^{+8.8}_{-12.5} $ & $ +0.52^{+0.42}_{-0.29} $ & \multicolumn{1}{c}{---} \\
    Theor & QCD sum rule (Case II)                                                      & $ -15.5^{+8.8}_{-12.5} $ & \multicolumn{1}{c}{---} & $ +0.18^{+0.14}_{-0.10} $ \\
    \hline
  \end{tabular}
\end{table}
\subsection{Theoretical determination}
\par
We have proposed that
both the mass difference of mirror nuclei $ \Delta E_{\urm{tot}} $ and
the neutron-skin thickness $ \Delta R_{np} $
depend linearly on $ s_0 $ in a way 
which is almost universal among the isospin symmetric part of the Skyrme interaction
if only the leading-order CSB interaction $ s_0 $ is considered with the fixed value of $ y_0 $~\cite{
  Naito2022Phys.Rev.C105_L021304}.
Here, we take $ \Delta E_{\urm{tot}} $ as an example.
The $ \Delta E_{\urm{tot}} $ with arbitrary $ s_0 $ can be parametrized as
\begin{equation}
  \Delta E_{\urm{tot}} \left( s_0 \right)
  =
  a s_0
  +
  \Delta E_{\urm{tot}}^{\urm{w/o CSB}},
\end{equation}
where $ \Delta E_{\urm{tot}}^{\urm{w/o CSB}} $ is the mass difference calculated without the CSB interaction.
The value of $ a $ is almost universal among Skyrme EDFs,
and hence the averaged value $ \overline{a} $ is almost equal to each $ a $.
Once \textit{ab initio} calculations provide $ \Delta E_{\urm{tot}} $ with and without the CSB interaction,
$ s_0 $ can be determined by
\begin{equation}
  s_0
  =
  \frac{\Delta E_{\urm{tot}}^{\urm{w/ CSB}} - \Delta E_{\urm{tot}}^{\urm{w/o CSB}}}{\overline{a}}.
\end{equation}
\par
Reference~\cite{
  Novario2023Phys.Rev.Lett.130_032501}
provides $ \Delta E_{\urm{tot}} $ of $ \nuc{Ca}{48}{} $ and $ \nuc{Ni}{48}{} $ with and without the CSB contribution
calculated by the coupled cluster (CC) approach with the $ \text{N}^2 \text{LO}_{\text{GO}} $ chiral interaction
without the Coulomb interaction,
whose value is
$ \Delta E_{\urm{tot}} = 0.72 \pm 1.10 \, \mathrm{MeV} $ for the $ \text{N}^2 \text{LO}_{\text{GO}} $ (394) interaction
and
$ \Delta E_{\urm{tot}} = 0.86 \pm 4.85 \, \mathrm{MeV} $ for the $ \text{N}^2 \text{LO}_{\text{GO}} $ (450) one.
Note that the mass difference $ \Delta E_{\urm{tot}} $ without the Coulomb interaction originating almost only from the CSB interaction
since the CIB interaction rarely contributes to $ \Delta E_{\urm{tot}} $~\cite{
  Naito2023Phys.Rev.C107_064302}.
Therefore, $ \Delta E_{\urm{tot}}^{\urm{w/o CSB}} $ is zero and
the aforementioned value can be used for $ \Delta E_{\urm{tot}}^{\urm{w/ CSB}} $.
Since the averaged slope is $ -0.3399 \pm 0.0046 \, \mathrm{fm}^{-3} $~\cite{
  Naito2022Phys.Rev.C105_L021304},
the obtained values of $ s_0 $ are
$ -2.1 \pm 3.2 \, \mathrm{MeV} \, \mathrm{fm}^3 $
and
$ -2.5 \pm 14.3 \, \mathrm{MeV} \, \mathrm{fm}^3 $
obtained, respectively, by 
the $ \text{N}^2 \text{LO}_{\text{GO}} $ (394)
and
the $ \text{N}^2 \text{LO}_{\text{GO}} $ (450) interactions.
\par
The mass difference $ \Delta E_{\urm{tot}} $ of $ \nuc{Be}{10}{} $-$ \nuc{C}{10}{} $ calculated by the Green's function Monte Carlo (GFMC) calculation with the Argonne v18 (AV18)~\cite{
  Wiringa1995Phys.Rev.C51_38}
and Urbana X (UX)~\cite{
  Wiringa2014Phys.Rev.C89_024305}
interactions is also available~\cite{
  Wiringa_}.
Using the contributions originates from the mass difference of a proton and a neutron
and
the CSB interaction (the 18th term of the AV18 interaction),
the total CSB contribution to $ E_{\urm{tot}} $
for $ \nuc{Be}{10}{} $ and $ \nuc{C}{10}{} $ 
are estimated as 
$ -0.0932 (6) $ and $ 0.0918 (6) \, \mathrm{MeV} $, respectively.
Since the averaged slope is $ -0.05769 \pm 0.00147 \, \mathrm{fm}^{-3} $,
the obtained $ s_0 $ is
$ -3.207 \pm 0.086 \, \mathrm{MeV} \, \mathrm{fm}^3 $.
\par
Another method to pin down the CSB strength is based on the sum rule in QCD~\cite{
  Sagawa:2023itk}.
The QCD sum rules relate the proton-neutron mass difference in symmetric nuclear matter
to the in-medium quark condensate associated with the partial restoration of chiral symmetry. 
Using this proton-neutron mass difference,
the mass difference of mirror nuclei can be calculated with the local density approximation,
whose density dependence is, indeed, the same as Eq.~\eqref{eq:CSB_LDA}.
Hence, not only $ \tilde{s}_0 $ but also $ \tilde{s}_1 + 3 \tilde{s}_2 $ can be determined.
\par
The strengths obtained by these methods are summarized in Table~\ref{tab:CSB}.
The obtained $ \tilde{s}_0 $ based on \textit{ab initio} methods are around $ -5 \, \mathrm{MeV} \, \mathrm{fm}^3 $,
which is remarkably smaller than the phenomenological determination.
Especially, $ \tilde{s}_0 $ of the SLy4-ISB, SkM*-ISB, and $ \text{SV}_{\text{T}} $-ISB interactions are also determined by using the mass difference,
which is the same as \textit{ab initio} determination.
Even though the same quantities are used to determine $ \tilde{s}_0 $,
the strength determined by \textit{ab initio} calculation is $ 1/5 $ of the phenomenological value.
The possible reasons for this inconsistency may be the following:
experimental data, in principle, include many correlations beyond mean-field that are hard to capture in current EDFs like those considered in this study;
the $ \nuc{Be}{10}{} $-$ \nuc{C}{10}{} $ pair may be too light to compare with the mean-field or DFT calculation or they are deformed;
the chiral interaction may not be accurate enough or converged well~\cite{
  Machleidt:2023jws}
to discuss the small contribution such as the ISB terms.
\par
As for the next-leading-order CSB interaction,
$ \tilde{s}_0 $ obtained by the QCD sum rule is not so significantly different from the phenomenological values,
while $ \tilde{s}_1 $ and $ \tilde{s}_2 $ are quite small.
This is remarkably in contrast to the next-leading-order $ \text{SV}_{\text{T}} $-ISB interaction.
As seen in Eq.~\eqref{eq:CSB_LDA}, the sign of $ \tilde{s}_0 $-term and those of $ \tilde{s}_1 $- or $ \tilde{s}_2 $-term are opposite;
accordingly, their effects on the energy density $ \ca{E}_{\urm{CSB}} $ are opposite to each other.
Therefore, it may be difficult to pin down all $ \tilde{s}_0 $, $ \tilde{s}_1 $, and $ \tilde{s}_2 $ at the same time from experimental data.
%
%
%
\section{Summary}
\par
In this proceeding, we discussed the strength of the charge symmetry breaking term of an effective nuclear interaction or an energy density functional.
There is an inconsistency between the strength determined phenomenologically and theoretically.
It is discussed that the isospin symmetry breaking terms of nuclear interaction themselves are small,
but they sometimes give non-negligible or even significant contributions~\cite{
  Naito2022Phys.Rev.C106_L061306,
  Naito2023Phys.Rev.C107_064302}.
Therefore, it is indispensable to solve such an inconsistency and to pin down the strengths properly.
%
%
\acknowledgments
The authors thank
Andreas Ekstr\"{o}m,
Stefano Gandolfi,
and Robert B.~Wiringa for fruitful discussions.
T.N.~and H.L.~would like to thank the RIKEN iTHEMS program
and the RIKEN Pioneering Project: Evolution of Matter in the Universe.
T.N.~acknowledges
the RIKEN Special Postdoctoral Researcher Program.
G.C.~gratefully acknowledges the support and hospitality of YITP, Kyoto University, where part of this work has been carried out.
This work is partly supported by 
the JSPS Grant-in-Aid under Grant Nos.~18K13549, 19K03858, 20H05648, 22K20372, 23H04526, 23H01845, and 23K03426.
The numerical calculations were performed on cluster computers at the RIKEN iTHEMS program.
%
%

%
\end{document}